\documentstyle[12pt]{article}
\begin{document}

\title{Microlensing of disk sources}
\author{S. Mollerach and E. Roulet\thanks{Also INFN, Sezione di Trieste} \\
{\it International School for Advanced Studies, SISSA}\\
Via Beirut 2, I-34013, Trieste, Italy}
\date{}
\maketitle
\vfill
\begin{abstract}
We analyse the effects on the predictions for the microlensing
searches toward the Galactic bulge coming from the fact that not all
the stars monitored belong  to the bulge itself, but that a
non--negligible fraction of them actually are in the Galactic disk.
 The different distribution and motions of these disk stars make their
associated microlensing rates and event duration distributions to be
quite different from those of the bulge stars. We discuss the
uncertainties in these predictions associated to the modeling of the
Galactic components and the main implications resulting from the
inclusion of this second source population.

\end{abstract}
\noindent {\it Subject headings:} Galaxy: structure --- gravitational lensing
\vfill
\clearpage

Three experiments, MACHO (Alcock et al. 1995), OGLE (Udalski et
al. 1994) and DUO (Alard et al. 1995),  are actively searching for
microlensing events in the direction of the Galactic bulge, and their
observations are already providing important information on the
morphology of the inner Galaxy. Essentially two lensing populations
are believed to be responsible for the events: the objects in the disk
 and those in the bulge itself (Griest et al. 1991, Paczy\'nski 1991,
Giudice et al. 1994, Kiraga \& Paczy\'nski 1994). The
large rates observed are most probably the result of the barred shape
of the bulge (Kiraga \& Paczy\'nski 1994, Paczy\'nski et al. 1994,
Zhao et al. 1995, Han \& Gould 1995), with the major bar axis making a
small angle with respect to the line of sight to the Galactic
centre. A disk close to maximal would also be
helpful in explaining the large rates observed.

Up to now, an assumption made in all theoretical analyses is that the source
population consists of the bulge stars alone. However, it is known
that, in the fields observed, a non--negligible fraction of the source
stars actually belong to the disk. For instance, Terndrup (1988)
estimated that 15\% of the red giant stars in Baade's Window, at 
$b=-3.9^\circ$ and $l=1^\circ$, belong to the disk. Probably this
fraction is even larger for main sequence stars in the same field, due
to the younger age of the disk. The fraction of disk stars should 
increase in fields at larger longitudes and similar latitudes, where
microlensing observations are also carried out. It is our purpose in
this letter to discuss the possible implications of this second source
population for the analyses of microlensing results.

Let us assume that, in a given field, a fraction $f$ of the monitored stars
belong to the disk, while $1-f$ is the fraction of bulge stars. We
denote $\tau_{ls}$  the 
optical depth to microlensing of disk or bulge sources
($s=D,B$) arising from disk or bulge lenses ($l=D,B$). Hence, the total
optical depth is just
\begin{equation}
\tau=f[\tau_{DD}+\tau_{BD}]+(1-f)[\tau_{DB}+\tau_{BB}].
\end{equation}
We recall that the optical depth is averaged over the source
distribution (Kiraga \& Paczy\'nski 1994)
\begin{equation}
\tau_{ls}={1\over N_s}\int_0^{D_{max}}dD_{os}{dn_s\over dD_{os}} 
\int_0^{D_{os}}dD_{ol}{4\pi G\over c^2D_{os}}\rho_l D_{ol}(D_{os}-D_{ol}),
\end{equation}
where $D_{os}$ and $D_{ol}$ are the observer distances to the source
and the lens respectively, $\rho_l$ is the
lens mass density, $dn_s/dD_{os}\propto \rho_s D_{os}^{2-2\beta}$
describes the number density profile of detectable sources along the 
line of sight (Kiraga \& Paczy\'nski 1994), and the normalization factor is 
$N_s=\int_0^{D_{max}}dD_{os}dn_s/dD_{os}$.
 The parameter $\beta$ arises because the
fraction of sources with luminosities larger than $L$ is assumed to
scale as $L^{-\beta}$. A reasonable range has been estimated to
be $\beta=1\pm 0.5$ in Baade's Window (Zhao et al. 1995), valid for
4~kpc$<D_{os}<12$~kpc. We will adopt in our discussion $\beta=1$,
using this value also for nearby sources in the disk
($D_{os}<4$~kpc), for which actually $\beta$ should be somewhat smaller,   
since we do not expect the main results to change significantly with
this last assumption. 

We will adopt for the disk distribution a double exponential profile
with constant scale height 325~pc (Bahcall 1986), scale length 3.5~kpc
and local density, in lensing objects, 
 $\rho_d=0.1 M_\odot$/pc$^3$, assuming hereafter that
the solar galactocentric distance is $R_0=8.5$~kpc. We will show results
for $D_{max}=12$~kpc and 6~kpc, this last value in order to illustrate
the possible effects of having a `hollow' disk, as could be suggested
by a certain deficit of disk stars beyond $D_{os}\simeq 3$~kpc
inferred from the color--magnitude diagram (CMD) 
in observations toward the bulge (Paczy\'nski et
al. 1994b).
 We note that
these observations may also suggest that the scale height is
smaller toward the centre (see also Kent et al. 1991). Finally,
measurements of the scale length provide values in the wide range from 1.8
to 6~kpc (Kent et al. 1991), further increasing the uncertainties associated
to disk models. However, the hollow and non--hollow disk models
considered 
already span a representative range of possible expected results.

\begin{table}
\begin{center}
\begin{tabular}{cccccccc} \hline\hline
  $ls$ &   $D_{max}$ & $\tau_{ls}$ & $\Gamma_{ls}$ &
$\langle T\rangle$ & $\tau_{ls}$ & $\Gamma_{ls}$ &
$\langle T\rangle$ \\
& [kpc] & \multicolumn{3}{c}{ Baade's Window} & \multicolumn{3}{c}{ $(b,l)=(3^\circ,10^\circ)$}\\
\hline
DD & 12 & 0.58 & 2.1 & 26 & 0.70 & 2.3 & 31\\
   & 6 & 0.16 & 0.4 & 44 & 0.19 & 0.5 & 44\\
BD & 12 & 0.77 & 3.4 & 20 & 0.62 & 2.3 & 25\\
   & 6 & 0.004 & 0.02 & 8 & 0.01 & 0.07 & 9 \\
DB & 12 & 0.94 & 3.5 & 25 & 0.74 & 2.8 & 23\\
   & 6 & 0.70 & 2.3 & 29 & 0.65 & 2.4 & 25\\
BB &  & 1.32 & 5.3 & 22 & 0.46 & 2.0 & 20\\
\hline
\end{tabular}
\end{center}
\caption{Contributions to the optical depth $\tau_{ls}$ ($\times
10^{-6}$), rates
$\Gamma_{ls}$ (in events/$10^6$~stars yr) and average event durations
$\langle T\rangle$ (in days), for lens and source
populations in the disk ($D$) or bar ($B$). For $\Gamma$ and $\langle
T\rangle$ we assume  
 lenses of mass $m=0.2 M_\odot$ and use the OGLE efficiency. The first
three columns are for Baade's Window, the last three for $l=+10^\circ$
and $|b|=3^\circ$.}
\end{table}

For the bulge stars, we will use the bar model of Dwek et
al. (1995), obtained from COBE--$DIRBE$ maps of the Bulge. 
The existence of a bar was initially suggested by several bulge 
observations, including an asymmetry in the  infrared surface
brightness distribution and in the star luminosities at positive and
negative longitudes,  and also to
explain the non--circular gas motions observed
 (Spergel 1992). 
We adopt a total
bar mass $M_{bar}=2\times 10^{10}M_\odot$, in the upper range of
different dynamical estimates (see however Blum 1995), as suggested by 
microlensing observations. The optical depths and rates due
to bar lenses are of course proportional to $M_{bar}$, 
while the event durations
are insensitive to it. The angle between the bar major axis and the
direction to the Galactic centre is taken to be $\alpha=20^\circ$, in
the middle of the range $\alpha=20^\circ\pm 10^\circ$ obtained by Dwek
et al.. We take the same velocity dispersion of bar objects
as in Han \& Gould (1995).

In the third column of Table~1 we give the predictions for the
different components of the optical depth toward Baade's
Window. Since the optical depth of disk sources,
$\tau_D\equiv\tau_{DD}+\tau_{BD}$, is always smaller than the one for
bulge sources, $\tau_B\equiv\tau_{DB}+\tau_{BB}$, taking into account
the disk sources will lead to a smaller theoretical prediction for the total
depth than in the case $f=0$. 
This in turn will imply an underestimate, as noted by Bennet et
al. (1994), of the required $\tau_B$, and hence of the inferred
 bar and/or disk total mass normalizations required.
For instance, if we assume $f=20$\% in Baade's Window, 
$\tau$ is a factor 0.92
smaller than $\tau_B$  if $D_{max}=12$~kpc is
considered, 
while it is a factor 
$\simeq 1-f=0.8$ smaller if $D_{max}=6$~kpc, since  $\tau_D$ turns out to be
very small in this case. 
We also see that $\tau_D$ is very sensitive to $D_{max}$,
but $\tau_B$ has a milder dependence on it which arises only 
through $\tau_{DB}$.

\begin{figure}
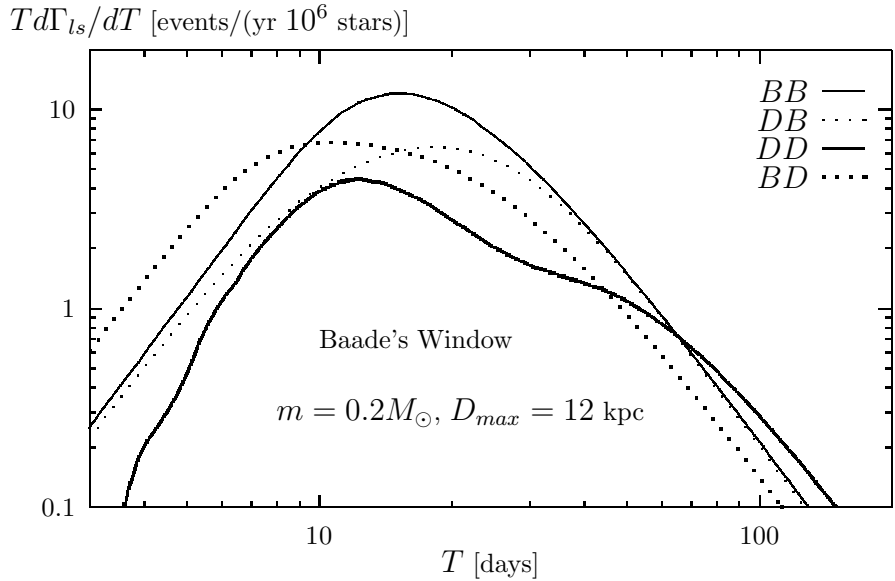

\begin{center}
\input dgdtbw212
\end{center}
\caption{Differential rate distribution in Baade's Window, assuming a
common lens mass $m=0.2M_\odot$ and a disk extending up to
$D_{max}=12$~kpc. The curves are labeled by $ls$, where $l,s=D,B$ for
objects in  the disk or in the bar respectively.}
\end{figure}

We turn now to discuss the rates and event duration distributions. Of
course, the rates have a similar decomposition as the optical depth in
eq.~(1). In
Fig.~1 we show the differential rate for the different components 
as a function of the event duration, for Baade's Window
observations. 
We  assume
$D_{max}=12$~kpc and take for definiteness 
all lenses to have a common mass $m=0.2
M_\odot$. 

A crucial difference between the disk and bulge sources,
besides their spatial distribution, is their motion. Since disk
objects have small velocity dispersion, here taken to be
$\sigma=30$~km/s\footnote{the local dispersion of disk stars is $\simeq
20$~km/s, but is larger toward the Galactic centre (see e.g. Lewis \&
Freeman 1989)},
and a similar global motion due to rotation as the observer's one, 
 $DD$ type events are expected to have particularly long durations when
$D_{os}<R_0$.  
These long duration events in $\Gamma_{DD}$, centered
around $40\div 80$~d, can clearly be identified in Fig.~1. 
There is a second contribution to $\Gamma_{DD}$ at
smaller times ($\sim 5\div 30$~d), due to events where the source star
lies at $D_{os}>R_0$, so that it  is moving in the opposite direction than
most of the disk lenses, which are at $D_{ol}<R_0$. 
We assumed the circular speed of the
disk to be everywhere 220~km/s, and we expect small modifications of
the results obtained due to the slower actual rotation in the very inner bulge.
It is apparent from Fig.~1 that the rates associated to disk stars are
relatively more important at longer event durations, since short duration
events are overwhelmingly due to $\Gamma_{BB}$ (recall that the
contributions to $\Gamma$ from disk and bulge sources are weighted by
$f$ and $1-f$ respectively, and that the observational efficiencies
are generally very small for $T<10$~d). 

\begin{figure}
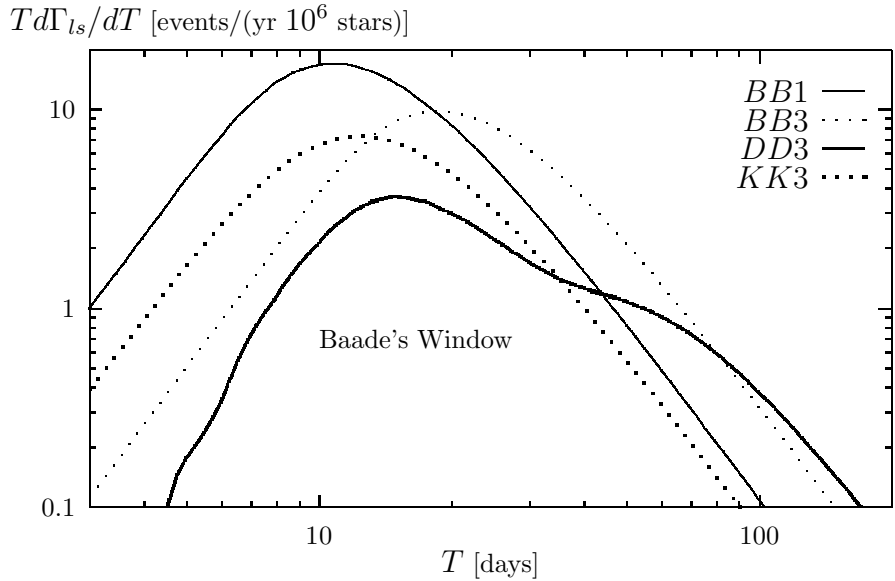

\begin{center}
\input dgdtbw3
\end{center}
\caption{As in Fig.~1, but for bar models with different lens masses,
$m=0.1M_\odot$ ($BB1$) and $m=0.3M_\odot$ ($BB3$) and for Kent's
axisymmetric model with $m=0.3M_\odot$ ($KK3)$.  We also show
the $DD$ distribution for $m=0.3M_\odot$.}
\end{figure}

In order to illustrate the dependence of the results in the
assumptions done for Fig.~1, we show in Figs.~2 and 3 some variations of
the assumed models.  Fig.~2 considers the effect of changing the lens
mass as well as the bulge model. There is no reason to expect that the
mass function of disk lenses will be similar to that of bulge
lenses. As an example to illustrate the implications that this may
have,  we show the predictions for $BB$ lensing for masses
$m=0.1M_\odot$ (curve $BB1$) and $0.3 M_\odot$ (curve $BB3$), 
and compare them to the prediction of $DD$ with $m=0.3M_\odot$ (curve
$DD3$).  If both masses are $0.3M_\odot$, we have the same situation
as in Fig.~1 slightly shifted to larger times (and with reduced
rates). However,  if the  bar lenses have a mass smaller than the disk
lenses,
the relative contribution to the differential rate 
coming from the disk sources is more
important at large times than before. Similar conclusions clearly
follow if the total bar mass is actually smaller than the assumed
$2\times 10^{10}M_\odot$. 
In the same way, 
if the triaxiality of the bulge is reduced, the contribution coming
from the bulge sources would be smaller than before.
For instance,  we
 show the predictions ($KK3$ curve) for bulge--bulge 
lensing adopting
Kent's axisymmetric model of the bulge (Kent 1992), obtained from
Spacelab data, and assuming $m=0.3M_\odot$. 
This model has a more centrally concentrated mass
distribution and larger transverse velocity motions than the bar
model, leading to
shorter duration events and smaller rates. Similar results are
obtained (Giudice et al. 1994, De R\'ujula et al 1995) with 
the spherically symmetric heavy spheroid model of
Caldwell and Ostriker, in which
the bulge is just the inner spheroid. Although these models fail to
account for the observed optical depth to the bulge, Fig.~2 shows that
any deviation of the bar model in the direction of the axisymmetric
ones would tend to enhance the $DD$ contribution at large event
durations.

Fig.~3 instead shows the effect of considering a `hollow' disk. We
depict, for $m=0.2M_\odot$ as in Fig.~1, the $DD$ and $DB$
contributions for $D_{max}=6$ and 12~kpc. Clearly the $BD$
contribution, not depicted, becomes negligible for $D_{max}=6$~kpc. We
see that the long duration tail ($T>30$~d) of the $DB$ rates is not
greatly affected, since it is mainly due to lenses at $D_{ol}<6$~kpc,
far from the rapidly moving bulge sources. Instead, the $DD$ rates
become suppressed, with the short duration events clearly
disappearing.

\begin{figure}
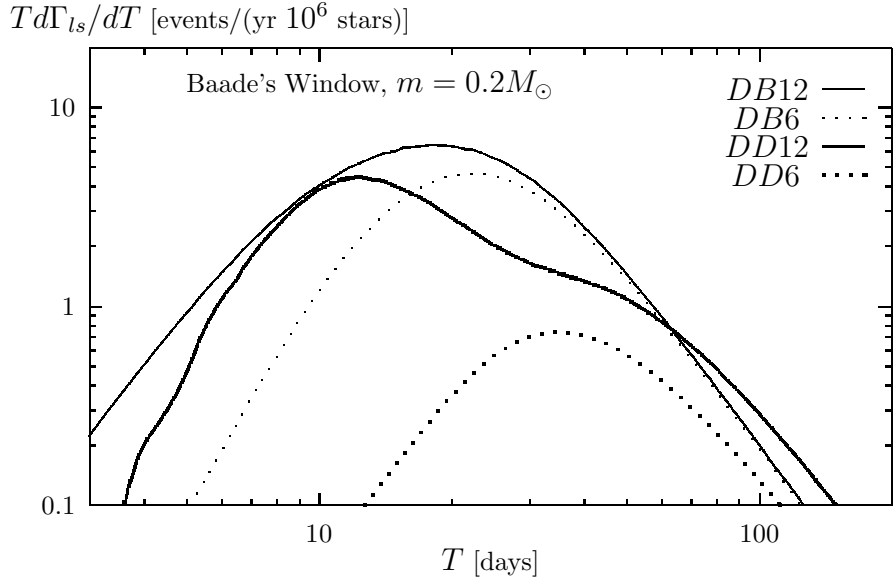

\begin{center}
\input dgdtbwd
\end{center}
\caption{Predictions for hollow ($D_{max}=6$~kpc) and non--hollow
($D_{max}=12$~kpc) disk models for Baade's Window.}
\end{figure}

In Fig.~4 we show the differential rate predictions at
$|b|=3^\circ$ and $l=\pm 10^\circ$, for $m=0.2M_\odot$ and
$D_{max}=12$~kpc as in Fig.~1. 
The $DD$ contribution is not significantly changed with respect
to the situation in Baade's Window, except that the distinction
between short duration events and long duration ones is less clear,
due to the different transverse motion of disk objects at larger
longitudes.
The $BB$ rates turn out to be less  important in these fields.
 The rates involving
the bar change significantly in the two fields, due to the inclination
of the bar, and make $\Gamma_{DB}$ quite important at negative
longitudes, for the assumed $D_{max}=12$~kpc value, while
$\Gamma_{BD}$ (not plotted) becomes quite small.
At positive longitudes, on the other hand, $\Gamma_{DB}$ is much
smaller, $\Gamma_{BD}$ is sizeable,  and 
 the $DD$ contribution to the rate is the largest one for 
long duration events. Since the
fraction of disk stars also increases with longitude, it becomes even
more important to take the effects of disk sources into consideration in the
microlensing searches in these fields.

\begin{figure}
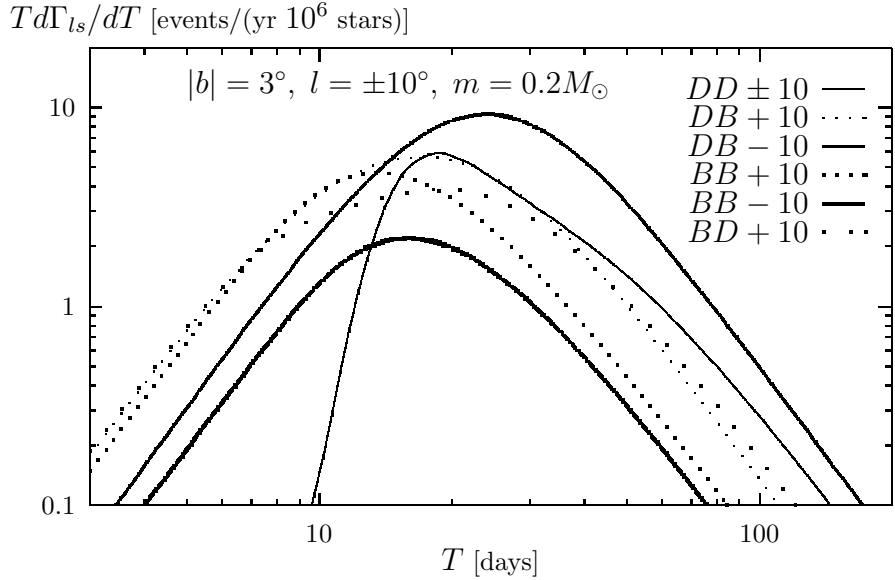

\begin{center}
\input dgdt310
\end{center}
\caption{Differential rates in fields at $|b|=3^\circ$ and
$l=\pm10^\circ$, assuming $m=0.2M_\odot$ and $D_{max}=12$~kpc.}
\end{figure}

The fact that $\Gamma_{BD}$ has the opposite behaviour than
$\Gamma_{DB}$ in fields at positive and negative longitudes 
has the effect of reducing the asymmetric signatures in the
microlensing maps of the bulge when disk source stars are taken into
account. For instance, we obtain, for $D_{max}=12$~kpc and
$|b|=3^\circ$, 
that $\tau(l=10^\circ)/\tau(l=-10^\circ)\simeq 1.5$ if $f=0$, while this
ratio is $\simeq 0.94$ if we take $f\simeq 0.4$ (0.6) for $l=10^\circ$
($-10^\circ$)\footnote{We note that a rough estimate of the angular
dependence of the fraction $f$, in clear windows, may be obtained as
$f/(1-f)=\kappa N_D/N_B$, where $\kappa$ is a constant chosen such
that $f\simeq 0.2$ at Baade's Window, and $N_D$ and $N_B$ are the
values of $N_s$ in eq. (2) computed for disk and bulge sources
respectively. This leads to $f=0.4$ (0.6) for $|b|=3^\circ$ and
$l=+10^\circ$ ($-10^\circ$).}. We note that 
 the asymmetry of the individual contributions $\tau_{ls}$ 
between fields at positive and negative
longitudes  depends sensitively on the bar inclination
assumed (and should disappear for $\alpha=0$ clearly), but due to the
above mentioned cancellations, it should be harder to get information
on $\alpha$ from the asymmetries in the microlensing maps. However,
the size of $\tau$ do depend on $\alpha$. 
 For instance, the bar--bar optical
depth at  Baade's Window 
 takes the values $\tau_{BB}(\alpha=30^\circ)= 0.97\times 10^{-6}$, while
$\tau_{BB}(\alpha=10^\circ)= 1.74\times 10^{-6}$, and hence the large
rates observed suggest that the inclination is small.

Going back to Table~1, we show in the last five columns the total
rates and average event durations, assuming $m=0.2M_\odot$ and 
the observational
efficiency of the OGLE experiment, for Baade's Window and for a field
at $|b|=3^\circ$ and $l=+10^\circ$, as well as the optical depth in
this last field. The average durations do not change significantly
among the two fields, and reflect the behaviour of the differential
rates just discussed in Figs.~1--4. 
From the total rates (including the efficiency), we may
estimate the fraction $F$ of events which result from disk sources, which
is just $F=f\Gamma_D/\Gamma$. For Baade's Window, taking $f=0.2$ we
obtain $F\simeq 0.13$ if $D_{max}=12$~kpc, while $F=0.01$ if
$D_{max}=6$~kpc. In the field at $|b|=3^\circ$ and $l=+10^\circ$ we
obtain instead, assuming for illustrative purposes  $f=0.4$, that
$F\simeq 0.39$ if $D_{max}=12$~kpc, while $F=0.08$ if
$D_{max}=6$~kpc. So, it is quite plausible that some of the events
among the almost 100 events observed up to now are  due to disk
sources. 
This fact is 
interesting also because it has been suggested that there is 
actually a possible excess of long
duration events with respect to the predictions made with
$\Gamma_B\equiv\Gamma_{DB}+\Gamma_{BB}$ alone (Han \& Gould 1995b). The long
duration events associated to disk sources may help in this respect,
since the total $d\Gamma/dT$ distribution has an enhanced tail at
large $T$.
Finally, the fact that the time distribution of events is different
when the disk sources are taken into account also affects the
determination of the mass functions of the disk and bulge lens
populations.

In the longer term, it may become possible to get, with enlarged
statistics, more information about the microlensing from the 
disk source population, considering
the lensing effects of stars in certain regions of the
 CMD. In particular, the bluer end of the main sequence
and the brightest red clump stars, should  in their
majority be foreground disk stars (Paczy\'nski et al. 1994b). 
In the same way, considering only the region of
the CMD where the bulge red clump stars lie, it is possible to
minimize the contamination of disk stars (Bennet et al. 1994), though
never eliminate it\footnote{It is interesting to note that this cut
seems already to lead to a larger optical depth than the standard cut
including all sources (Bennet et al. 1994).}.
Moreover, the study of line of sight velocity
dispersions of the lensed sources, and eventually of their proper
motions,  should also help to deduce their
identity (Terndrup et al. 1995). 
Stars in a bar should have particularly large line of sight dispersion
 while the dispersion of disk stars should be much smaller.

As a summary, we have considered the effects resulting from the fact
that a fraction of the sources observed in microlensing searches
toward the bulge belong to the disk. 
Since their optical depth is typically smaller than the one of bulge
sources, the required $\tau_B$ needs to be larger than the measured
optical depth, affecting then the model parameters inferred. This
effect is maximum if the disk is hollow (assuming the same $f$), but
is probably the only effect resulting from these disk models.
If the disk is not hollow, the rates due to disk stars are also
relevant. They are relatively more important for long duration events
and will then modify the shape of the event duration distribution. The
relative contribution from disk and bulge sources also depends on the
mass function distributions of the two lensing populations, on the
overall normalization of their densities and on the details of the bulge
model. The contribution from disk sources increases with increasing
longitudes, and can even become comparable to the one from bulge
sources in the extreme field considered ($l=10^\circ$,
$b=3^\circ$). For the barred model of the bulge, the longitude
dependence of the rates has the opposite behaviour for disk sources
than for bulge ones, reducing the overall asymmetries of the
microlensing maps of the bulge.

\clearpage
\bigskip

REFERENCES
\bigskip

\noindent Alard, C., Mao, S., \& Guibert, J. 1995, A\&A, submitted

\noindent  Alcock, C., et al. 1995, ApJ, 445, 133

\noindent  Bahcall, J. N. 1986, ARA\&A, 24, 577

\noindent  Bennet, D. P., et al. 1994, preprint

\noindent  Blum, R. D. 1995, ApJ, 444, L89

\noindent  De R\'ujula, A., Giudice, G., Mollerach, S., \& Roulet,
E. 1995, MNRAS, 275, 545

\noindent  Dwek, E., et al. 1995, ApJ, 445, 716

\noindent  Giudice, G., Mollerach, S., \& Roulet, E. 1994, Phys. Rev. D,
50, 2406

\noindent  Griest, K., et al. 1991, ApJ, 372, L79

\noindent  Han, C., \& Gould, A. 1995, ApJ, 447, 53

\noindent  Han, C., \& Gould, A. 1995b, ApJ, submitted

\noindent  Kent, S. M., Dame, T. M., \& Fazio, G. 1991, ApJ, 378, 131

\noindent  Kent, S. M. 1992, ApJ, 387, 181

\noindent  Kiraga, M., \& Paczy\'nski, B. 1994, ApJ, 430, L101

\noindent  Lewis, J. R., \& Freeman, K. C. 1989, AJ, 97, 139

\noindent  Paczy\'nski, B. 1991, ApJ, 371, L63

\noindent  Paczy\'nski, B., et al. 1994, ApJ, 435, L113

\noindent  Paczy\'nski, B., et al. 1994b, AJ, 107, 2060

\noindent  Spergel, D. N. 1992, in {\it The Center, Bulge, and Disk of the
Milky Way} (Kluwer, Dordrecht), 77

\noindent  Terndrup, D. M. 1988, AJ, 96, 884

\noindent Terndrup, D. M., Sadler, E. M., and Rich, R. M. 1995, AJ submitted

\noindent  Udalski, A., et al. 1994, Acta Astron., 44, 165

\noindent  Zhao, H., Spergel, D. N., \& Rich, R. M. 1995, ApJ, 440, L13

\end{document}